# On the Hourglass Model

Micah Beck
`mbeck@utk.edu`
Department of Electrical Engineering and Computer Science
University of Tennessee, Knoxville

## The Hourglass Model

The hourglass model of layered systems architecture [1] is a visual and conceptual representation of an approach to achieving a design that supports a great diversity of applications and admits a great diversity of implementations. At the center of the hourglass model is a distinguished layer in a stack of abstractions that is chosen as the sole means of accessing the resources of the system. This distinguished layer can be given implementations using components which are thought of as lying *below* it in the stack. The distinguished layer can be used to implement other services and applications that are thought of as lying *above* it. However, the components that lie above the distinguished layer cannot make direct access to the services that lie below it. The distinguished layer was called the "spanning layer" by Clark because it bridges the multiple implementation layers below it in the stack [2].

| Application 1 | Application 2 | Application 3 |
|---|---|---|
| Spanning Layer |||
| Implementation 1 | Implementation 2 | Implementation 3 |

The use of the hourglass model expresses the goal that the spanning layer should support many diverse applications and have many possible implementations. It also expresses the belief that restricting the functionality of the spanning layer is instrumental in achieving these goals. These elements of the model are combined visually in the form of an hourglass shape, with the "thin waist" of the hourglass representing the restricted spanning layer, and its large upper and lower bells representing the proliferation of applications and implementations, respectively.

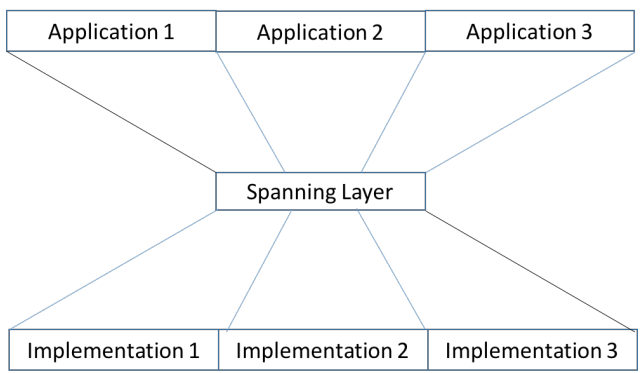

The hourglass model is a widely used as a means of describing the design of the Internet, and can be found in the introduction of many modern textbooks. It arguably also applies to the design of other successful spanning layers, notably the Unix operating system kernel interface, meaning the primitive system calls and the interactions between user processes and the kernel [3]. The impressive success of the Internet has led to a wider interest in using the hourglass model in other layered systems, with the goal of achieving similar results [4] [5]. However, application of the hourglass model has often led to controversy, perhaps in part because the language in which it has been expressed has been informal, and arguments for its validity have not been precise. Making a start on formalizing such an argument is the goal of this paper.

**The End-to-End Principle**
The most widely known and perhaps most controversial discussions relating to the hourglass model have been formulated as "The End-to-End Principle" or as a set of "end-to-end arguments" [1]. Sometimes the entire approach to thinking about system design is characterized as "End-to-End" without specification of a noun [6]. End-to-end arguments have had a huge influence on the thinking of system designers in the decades since the publication of the papers that named it, and some argue that the ideas that underlie those papers have even earlier origins.

The "End-to-End Principle" as sometimes presented can be paraphrased as follows: "In a layered architecture, any function should be located at the highest layer at which it can be correctly and completely implemented." [1] Focusing on the spanning, layer, this is taken to mean the functionality of the spanning layer should be minimized.

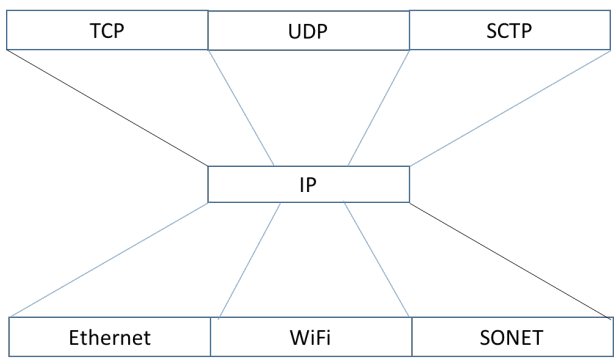

The term "end-to-end" derives from the fact that in the Internet architecture, the spanning layer is the network layer that implements transmission of datagrams from sender to receiver. The network layer is implemented at intermediate nodes, but is used by clients located at network endpoints. Thus, any function that is implemented above the network layer is implemented at network endpoints, in an "end-to-end" manner. In layered systems that are not physically laid out in this way, the term is anachronistic, but it is still used.

Attempts to augment the functionality of the Internet have often run into resistance that makes reference to "violating the End-to-End Principle" [7]. Proposers of any augmentation of the functionality of the intermediate node beyond simple datagram delivery have historically been treated to dire warnings that such services "will not scale" or that such an augmented network will cease to exhibit the benefits of the hourglass model.

Such arguments have been contentious, and have not led to a clear explanation of how a scalable distributed system could incorporate services beyond datagram delivery, in particular persistence of data (or storage) and processing (or computation). This has led some designers to doubt the validity of end-to-end arguments, while still seeking to achieve the goals of the hourglass model. It is very tempting to draw a the familiar hourglass shape and label the thin waist with a favorite interface depicted as the spanning layer, without having a clear argument as to why the implied results (many applications, many implementations) are to be expected.

This paper has grown out of my own efforts to understand and evaluate end-to-end arguments, starting by giving a formal structure within which to reason about layered architectures.

**A Note on this Draft**
I am distributing this paper in its current draft form in the hopes of soliciting comments and corrections from colleagues who share my interest in understanding layered design as a tool in the design of future systems in the light of historical experience. I have not completed the technical portions of Appendix A because the formal apparatus of program logic is substantial and I believe that the minimal properties that I need to justify are quite modest and ultimately intuitive. I have not worked in the area of formal program logic decades, and while model theory has served as a guide in my analysis of the hourglass I am not sure if it is necessary to use it in making a sound argument.

In this paper I am attempting to bring together 1. my long unused experience with the tools of programming logic with 2. my partial understanding of the history of some of the most powerful ideas and technologies of operating systems and networking and with 3. my own efforts to understand and develop new types of flexible globally scalable distributed systems. I believe that what I relate from personal or second-hand experience of the most important projects of the history computer systems is relevant to understanding the design tasks that were being undertaken, even if I have garbled or inaccurately attributed some of that history. I seek to interpret and extrapolate from the foundational work of the designers of Multics, Unix, and the Internet Protocol (see appendix B for some pictures) and more recent or ongoing efforts in the areas of Active/Programmable Networking, Grid & Cloud Computing, Network Virtualization and SDN.

I have been lucky enough to meet and work with some of the founders and leaders in the field of computer system design as teachers, mentors or colleagues, and they are too numerous to name here. All have contributed to the professional and intellectual environment in which I have lived for almost 40 years, and I hope that these efforts are seen as they are intended, as respectful homage to such prior work.  I ask the reader's indulgence in reading this draft, correcting mistakes or providing direction for its development, and giving constructive feedback on how or if the structure it describes can be a useful tool in the development of system design.

**Overview**
We begin by presenting an abstract framework for reasoning about layered architectures and spanning layers in particular. We assume the existence of a notion that one layer *implements* another, which we characterize as an "implements" relation between "layer specifications". We do not give formal definitions for layer specifications or the implements relation here, but provide a sketch in Appendix A, as the complete formalization is detailed and does not add substantially to the understanding of the argument. We then derive definitions of "possible implementations" and "possible applications" of a layer. Our account of the hourglass rests on two simple properties of these definitions, which we do not prove here, but give intuitive arguments for. Formal proof of these arguments require a formal definition of the "implements" relation, and again a sketch is given in Appendix A. Replacing the sketches given in the appendix with a fully formal treatment and proofs is an ongoing project, but it is hoped that readers familiar with programming logic will find the argument convincing nonetheless.

These definitions and the properties that we infer provide a framework for characterizing a spanning layer in terms of the multiplicity of its applications and its implementations, and the relationship between these and the minimality of the spanning layer. It is within this framework that we ask how the goals of the hourglass model can be represented, and what the relationship is between the minimality of the spanning layer and achieving those goals.

We then use this analysis to argue for the validity of a principle that is closely related to end-to-end arguments, although its statement is somewhat more general, and makes reference to some terms (deployment scalability, as well as simplicity, genericness, generality and resource limitation) that we have not yet defined.

> **The Deployment Scalability Tradeoff:** There is an inherent tradeoff between the deployment scalability of a specification against the degree to which that specification is simple, generic, general and resource limited.

Finally, we argue that this tradeoff is a useful way of understanding some formulations of "The End-to-End Principle" and some "end-to-end arguments".

## 1. Definitions

1.1 Service Specifications and the Implements Relation

1.1.1 Definition: A *service specification* to be a formal description of the syntax and necessary properties of a programming interface (API).

A service specification S is an API: it specifies the behavior of certain program elements (functions or subprograms) through statements expressed in a programming logic. For instance, these might be such statements:

- $\forall\, A, B \in \mathbb{Z}[\, (A + 1) + B\; =\; (A + B) + 1\,]$
- $\forall\, X, Y \in \mathbb{N}[\, \{X > 0\}\, Y := X * X\, \{Y > X\}\,]$

In formal terms a service specification is a theory of the programming logic. We denote by $\Sigma$ the set of all such specifications expressed in the language of the specific logic.

1.1.2 Definition: A specification S1 is *weaker* than another specification S2 iff $S2 \vdash S1$. S1 is *strictly weaker* than S2 if it S1 is weaker than S2 but S2 is not weaker than S1.

1.3 In appendix A we define an *implements* relation $S \prec_P T$ between two service specifications S and T and a program P.

The implements relation is intended to be analogous to the "reduces to" relation of structural complexity theory. Less formally, we say that "In a model where API S is correctly instantiated, the program P correctly implements API T in terms of the instantiation of S."

1.2.1 Lemma: If S1 is weaker than S2 and $S1 \prec_P T$ then $S2 \prec_P T$
See Appendix A.

1.2.2 Lemma: If S1 is weaker than S2 and $T \prec_P S2$ then $T \prec_P S1$
See Appendix A.

1.3 Pre- and Postimages
We will express our formal analogs to scalability in terms of how large the sets of models are that can implement or can be implemented using a specification. We define the pre- and post-images of a specification under implementation as follows:

1.3.1 $pre_\Pi(S) = \{T \mid \exists P \in \Pi\, [\, T \prec_P S\, ]\}$
1.3.2 $post_\Pi(S) = \{T \mid \exists P \in \Pi\, [\, S \prec_P T\, ]\}$

The definitions are relative to the set of programs $\Pi$ that are considered as possible implementations of one layer in terms of another. We do not specify this set, because we know of no formal description of all "acceptable implementations" of one layer in terms of another. This is certainly a limited class, and is in fact finite since programs that are too large are considered unwieldy from a software engineering point of view. This class also changes over time, as hardware and software technology changes the set of capabilities that are available as implementation tools.

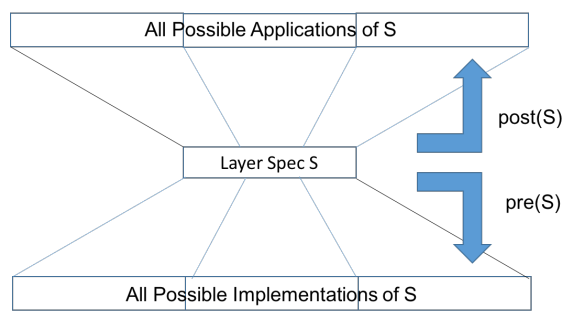

In representing elements such as the class $\Pi$ in our model as external parameters we acknowledge that our formalization only describes particular aspects of actual layered systems, and does not capture the entire structure. Hopefully our limited formal treatment can shed some light on how these aspects affect the part that we do analyze, and help give structure to the overall design process.

1.4 Using Pre- and Postimages As Analytical Tools
Reference to the hourglass model is sometimes conflated with the idea of the spanning layer as a standard that is enforced by some external means such as legal regulation or as a voluntary condition of membership in some community. However, we can use pre- and postimages of the implements relation as tools to analyze a layered system independently of any such application to the definition of standards.

If we choose any set of services at one level of a layered system, we can ask what the design consequences would be if it were adopted as the spanning layer of a hypothetical system. Adoption as a spanning layer means that no other services would be available at that layer, and that any participant in the system would have to use it as the sole means of accessing the services and resources of lower layers.

Viewed in this way, the preimage of the implements relation formally denotes all possible implementations of the prospective spanning layer and the post image denotes all possible applications. I use the term "denotes" because it is not necessarily a useful tool in actually enumerating these sets, since the parameter $\Pi$ has no formal specification and the question of whether a particular program P is in $\Pi$ is in general not computationally decidable. However, the question of what programs lies within the set we denote $\Pi$ has been the subject of much discussion in specific circumstances (notably in the acceptability of Internet protocols and applications), and having a formal characterization of how they relate to the specification of the spanning layer may be useful.

Taking this "descriptive" view of the hourglass allows us to use it as an analytical or predictive tool to understand the impact on communities of adopting particular interfaces as standards, be they de facto or de juris. Making the distinction between the use of the hourglass as a descriptive tool or as a means of justifying a standard also explains how many hourglasses can be understood as coexisting even within the same layered system. Every prospective spanning layer has an associated hourglass, irrespective of whether any of them are identified as standards. The impact of identifying a spanning layer as a standard is on the community that accepts that standard or has it imposed on them.

**2. The Hourglass Properties**
This theorem is central to our understanding of the hourglass model.

2.1 Theorem: If a specification S1 is weaker than another specification S2, then
   1. $post_\Pi(S1) \subseteq post_\Pi(S2)$, and
   2. $pre_\Pi(S1) \supseteq pre_\Pi(S2)$.

Proof:
1. By definition, if T$\in post_\Pi(S1)$, then
   - $\exists p \in \Pi \ [S1 \prec_p T]$, so by Lemma 1.2.1
   - $S2 \prec_p T$, thus
   - $T \in post_\Pi(S2)$.

2. By definition, if T$\in pre_\Pi(S2)$, then
   - $\exists p \in \Pi \ [T\ S2]$, so by Lemma 1.2.1
   - T $\prec_p$ S1, so
   - $T \in pre_\Pi(S1)$.

The hourglass properties tell us that a weaker layer specification has fewer applications but more implementations. While the latter conclusion corresponds to the intuition that underlies the idea behind the use of the hourglass model, the former may seem to contradict the value of the hourglass.

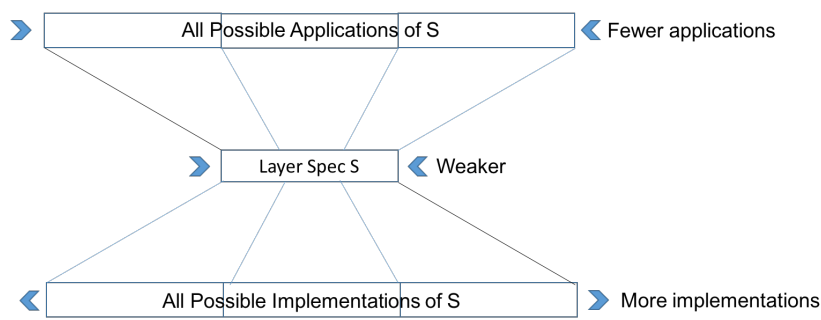

I note here that the hourglass properties are a theorem which has been proven, once complete proofs are given for Lemmas 1.2.1 and 1.2.2. It may be arguable whether the formal model given here fits a particular concrete scenario closely enough to be relevant. But within the confines of our model, the theorem will always hold. Thus, while a particular design may ignore the implications of this theorem, it cannot be "violated". Ignoring the implications of the theorem may or may not result in a less scalable design, but it certainly eliminates one possible means of achieving the goals of the hourglass model: many applications and implemenations.

### 3. Necessary Applications and Sufficient Specifications

The implication of the hourglass property on possible applications is that that a weaker spanning layer does not lead to more applications but to fewer. Thus, if weakness of the spanning layer is our tool for increasing possible implementations, we must introduce some countervailing element into the model to ensure that it is in fact possible to implement all necessary applications.

We model the design goal that it must be possible to implement certain applications by introducing the set of necessary applications as another external parameter $N \subseteq \Sigma$.

3.1 Definition: A specification S as *sufficient* to implement all necessary applications iff $N \subseteq post_\Pi(S)$.

This definition makes it clear that the design goals of application sufficiency and implementation richness are in tension. A spanning layer must be strong enough to implement all necessary applications but the stronger it is the fewer implementations are possible. We introduce the notion of *minimal sufficiency* as a means to balance the two

3.2 Definition: A specification to be *minimally sufficient for N* iff it is sufficient for N but there is no strictly weaker S' which is sufficient for N.

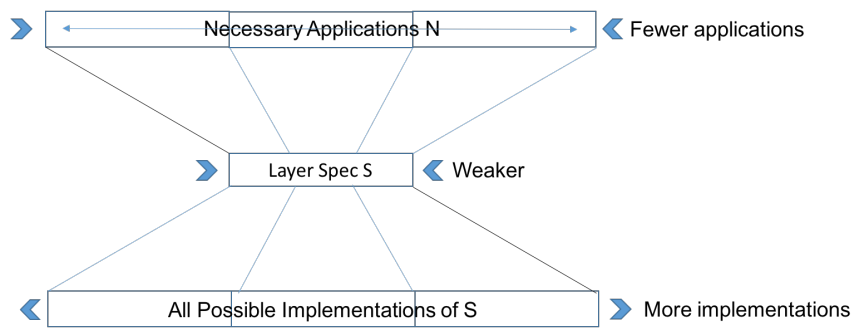

Thus the tension between application richness and implementation richness is achieved by specifying the former as a goal and then seeking a spanning layer that is as weak as possible to maximize the latter. This means that the choice of necessary applications N is in fact the most directly consequential element in the process of defining a spanning layer that meets the goals of the hourglass model.

Note the implication that the trade-off between the weakness of the spanning layer and its sufficiency is unavoidable. It predicts that, unless the effect is so small as to be negligible or overcome by other factors, attempts to achieve both may display inherent limitations. An analogy would be attempting to build a see-saw that allows both ends to be elevated simultaneously, or to simultaneously measure both velocity and position with unbounded accuracy. If this analysis is both accurate and relevant, then moving past it (sometimes referred to as "not adhering to the End-to-End Principle") may not be an attainable goal.

## 4. Genericness

While the choice of the particular set of necessary applications N is outside of the formal structure being presented in this paper, we can further characterize some implications of this choice. There is a great incentive for community members to see particular applications included in N, because if they are not included then those applications may not be implementable on top of the spanning layer. Users of applications lying outside of N may need to make use of services not included in the spanning layer, and if they do cannot be part of the community defined by it.

Let us assume that is a value metric $v: \mathcal{P}(\Sigma) \to \mathbb{R}$ is defined on sets of specifications. Then one characterization of the choice of a spanning layer S is whether it can be weakened without reducing the value of the set of possible applications. This leads to the following definition which is related to minimal sufficiency but incorporates the value metric. It specifies that there is no weakening of the specification which has a set of applications of greater value.

4.1 We define a specification to be *minimally sufficient for N relative to v* iff it is sufficient for N and if there is no S' strictly weaker than S for which $v(post_\Pi(S') \cap N) > v(N)$.

The value function v allows us to bring in unspecified considerations other than sufficiency as goals in choosing a spanning layer in order to maximize value. A good choice of spanning layer should not have a weakening which results in a set of applications with greater value.

But if weakness of the spanning layer is of primary importance, then it might be acceptable to accept a weakening which results in a set of applications of lower value, as long as it is not too much lower. That idea gives rise an even stronger condition.

4.2 Given $\mathcal{E} \in \mathbb{R}$ we define a specification to be $\mathcal{E}$-minimally sufficient for N relative to v iff it is sufficient for N and if there is no S' strictly weaker than S for which $v(post_\Pi(S') \cap N) - v(N) < \mathcal{E}$.

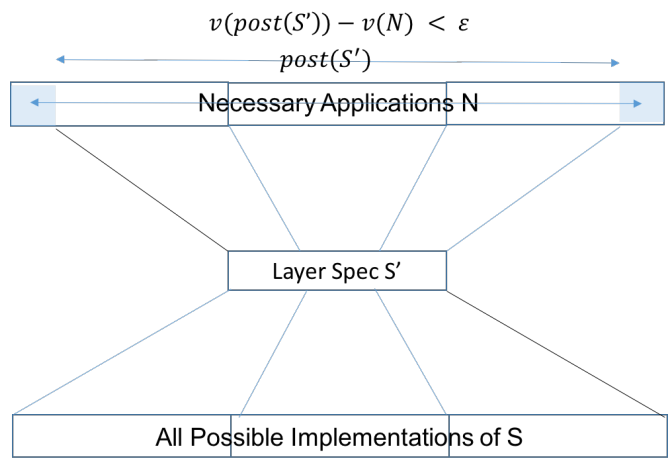

While this definition may seem complicated, involving no fewer than four external parameters (∏, N, v, and $\mathcal{E}$), it allows us to give a definition to the notion of *genericness* that has been an important part of the discussion of the Internet spanning layer.

Note that there is no guarantee that a specification that is $\mathcal{E}$-minimally sufficient for N relative to V exists for specific values of the external parameters. If the goal is to craft a generic spanning, some attention must be paid to these choices. Some necessary characteristics may be familiar from mathematics, corresponding to topological properties such as compactness. Since ∏, v, and $\mathcal{E}$ may be seen as fixed, this means that the requirements for the existence of an appropriate spanning layer may ned to be taken into consideration in making a choosing the set of necessary applications N.

In the context of the Internet, genericness is taken to mean that the spanning layer does not include features that are not of sufficient value to the entire community of prospective application users to justify the cost to the entire community of implementing them. In the context of the current discussion, the only such disadvantage that we can express is the need for a stronger spanning layer, resulting in fewer possible applications. Other possible notions of cost lie outside our formalism, including less simplicity, generality and greater resource consumption, as will be discussed in a later section.

4.3 Definition: A specification S to be *generic for N* iff it is $\mathcal{E}$-minimally sufficient for N relative to v for *an acceptable value of $\mathcal{E}$* where *v is a value function that models value to the entire community of users*.

Less formally, this definition means that a spanning layer is generic if there is no weakening of it which does not unduly reduce the value of the set of possible applications to the entire community of users. At the risk of repetitiveness, I point out again that the choices of $\prod, \mathcal{E}$ and v are undefined, and indicate the source of substantial potential for disagreement regarding the value of particular applications and the importance of minimality.

Alternative definitions of genericness do suggest themselves, some of them being even more complex. For instance, we could place a value metric on sets of possible implementation and then seek a spanning layer for which the slope of the trade-off between value of the set of applications and the value of increased implementations is sufficiently steep (the marginal cost of more applications is prohibitive relative to the value of the applications ruled out). I will eschew the further discussion of such complex constructions, but I do note that the of the definition of genericness is less intuitive than some other components of this formal model, and this may indicate that it may can benefit from further refinement.

**5. Applying the Hourglass to the Design of the Spanning Layer**
Our analysis of the implications of the choice of spanning layer to application and implementation richness gives us a tools for making such a choice when considering the design of a layered system.  If we agree on the limits of possible implementations (choice of $\prod$), the set of necessary applications (choice of N), our evaluation of the value of different sets of applications (choice of v and $\mathcal{E}$), then we can maximize the possible implementations of our spanning layer by choosing one that is generic for N.

This analysis does not tell us how to design such a spanning layer, but it does give us an account of the external factors that go into such a design, and how they interact to determine a desirable solution. Hopefully, discussion of the relative merits of those choices and of whether particular possible layers meet the definitions given above can be more structured than less formal discussions that make reference to no such model.

However, while this formal discussion may have become quite complex in its attempt to capture notions such as genericness that have been formulated and debated less formally, it is also incomplete in that it leaves out many considerations that have been considered part of the hourglass model but which do not have to do with the logical strength or weakness of the specification of the spanning layer. I will now attempt to relate some of those additional consideration to the formal model we have developed so far, in an attempt to more fully account for prior discussions of the hourglass model, including end-to-end arguments, by introducing additional attributes of the spanning layer: *simplicity, generality* and *resource limitation*.

### 6. Other Properties: Simplicity, Generality and Resource Limitation

The logical strength or weakness of the spanning layer is an appealing interpretation of the "thinness" of the spanning layer at the waist of the hourglass model largely because it yields to formalization using the tools of program logic. While this may account for some of the intention of prior references to the hourglass, it clearly does not capture it entirely, since other factors determine the value of a layer as a potential community standard [1].

#### 6.1 Simplicity

A requirement that is commonly given for the thin waist of the hourglass is that it must be *simple*. While logical weakness may be thought of as one aspect of simplicity, it clearly does not capture the entire concept. For example, one important aspect of simplicity that is not captured by logical weakness is orthogonality. In a service interface, orthogonality means that there is one way of implementing any function. Redundant features do not increase the strength of an interface but they do make it more complex. Software engineers understand the value of orthogonality in the design of interfaces and are more likely to accept a design that has this form of simplicity as a community standard, but it is not accounted for in our formal discussion.

We understand simplicity as an important aspect of the acceptability of the spanning layer as a tool to be used by humans and in other contexts where resources may be limited or other factors may affect its adoption. If software engineering metrics or other formalisms that can capture these aspects of the design, then they could be incorporated into a more complete version of our model.

#### 6.2 Generality

One unsettling aspect of this analysis of the hourglass is that it does not account for the incredible diversity of applications that are supported by the two most successful examples of this design approach: the Internet Protocol and tbe Unix kernel interface. Our analysis implies that logical weakness of the spanning layer does not contribute to the diversity of applications, and in fact acts against it.

So what accounts for the diversity we see in practice? It is often observed that the diversity of applications supported by the Internet far outstrips those forseen by its original designers. Thus we cannot say that the choice of necessary applications that went into the design directly determined the necessary strength of the spanning layer. (Perhaps the original designers are being modest, or had an implicit understanding of the eventual destiny of the network they were designing, but for the purposes of this discussion we will take them at their word.)

My belief is that the power of both of these systems is related to orthogonality. Rather than crafting a spanning layer to directly support the apparent needs of the target applications they were considering, the designers crafted a set of orthogonal primitives such that all the target applications lay within the space of applications generated by them. The consequence of this approach is that a well-crafted set of primitives is both an efficient strategy for implementing the set of target applications and also generates a highly diverse set of other applications they have not even been considered yet. In terms of our model, the design of the spanning layer S yielded an very high value of $v(post_\Pi(S))$. While the condition of sufficiency for a set of necessary applications is a more-or-less verifiable condition $N \subseteq post_\Pi(S)$, the value of all possible applications of a given spanning layer is much harder to evaluate, and designing a layer which tends to maximize it is still an art.

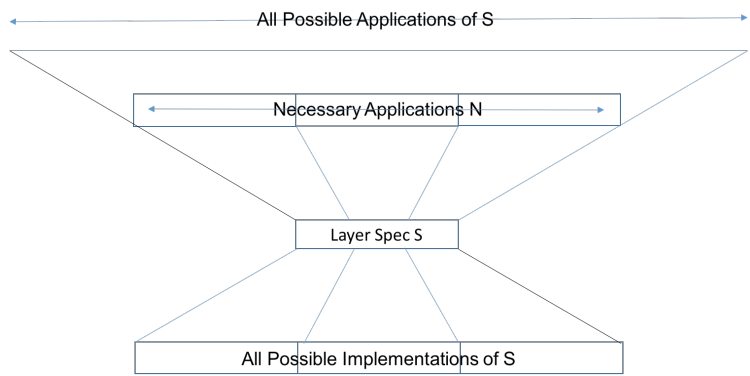

Neither the Internet nor Unix would have had the impact they have achieved without generality. One clue as to the origin of this design imperative within both the designers of the Internet Protocol and the Unix kernel interface may lie in a historical fact: Ken Thompson, Dennis Richie, Gerald Saltzer, David Clark and David Reed all participated in the Multics project, as did many of the prominent systems researchers of their generation [8]. Multics was an operating system project known for its many innovative features and which had substantial success in reaching many of its technical goals, but which was also known for extreme complexity and lack of orthogonality.

Multics is a classic example of a system that achieved its functionality goals but did not scale well. It is at least a workable hypothesis (which some unverified quotes attributed to Thompson substantiate) that this component of the designs that resulted in the most successful and scalable infrastructure interfaces in the history of computer systems, the Internet and the Unix operating system [3], were at least in part informed by the negative example of Multics, particularly in the areas of simplicity and generality. Further discuss of this hypothesis would require deeper delving into the history of computer systems.

**6.3 Resource Limitation**

The spanning layer provides an abstraction of the resources used in its implementation, preventing them from being accessed directly by applications. A such, it also defines the mechanism by which those resources are shared by applications and among users. In some communities, the modes of sharing are open, with few restrictions intended to ensure fairness among users (eg resource quotas). Such openness is one way of enabling the spanning layer to be logically weak (eg not implementing detailed dynamic authorization of user requests).

One way of managing more open modes of resource sharing is to limit the resources used by any individual service request, requiring large allocations of resources to be fragmented. Such fragmentation allows for more fluidity in the allocation of resources (eg storage allocations), with competition between users occurring on a finer scale and enabling subsequent services requests acting on allocated resources (eg movement of data between storage allocations) to also be limited in their use of resources.

Resource limitation means that use of the specification by an acceptable program will not result in overtaxing the resources of the platform on which it is implemented. In other words, the thin waist of the hourglass is also a thin straw through which applications can draw upon the unprotected resources that are available in the lower layers of the stack. Resource limitation does not have a direct impact on the logical strength or weakness of the spanning layer, but it can affect the ability of the system to function in environments where there the demand for resources locally or transiently exceeds the capacity of the system.

**7. The Deployment Scalability Tradeoff**

We have defined a model of a layered system of specifications and proved some properties relating the logical strength or weakness of one layer to the sets of possible applications and implementations. We then introduced the notion of a set of necessary applications as a design requirement of a spanning layer and then defined some characteristics that seek to characterize the fitness of a specification in meeting that requirement.

To augment this formal development, we have introduced three other ways of characterizing the "thinness" of a spanning layer: simplicity, generality and resource limitation. Together with our constructed notion of genericness, we now seek to account for the idea that a system built on the hourglass model is well adapted to finding success in the form of widespread adoption. We begin by giving a definition to this admittedly imprecise notion of success.

7.1 We define *deployment scalability* as widespread acceptance, implementation and use of a service specification.

Deployment scalability is a problematic choice of goal because we have no clear way to specify whether or not it has been achieved. But as we are attempting to account for informal arguments, we may have to live with that. Our formal model depends on parameters that a community may have trouble agreeing on: acceptable programs, a set of goal states, a vague "value function" and the acceptable tradeoff between increased implementations and supported applications. Then finally, we have added in three unformalized notions that we believe also influence the fitness of a spanning layer to achieve deployment scalability.

Undaunted, we offer a principle that is not a hard-and-fast rule but a tradeoff between these problematic elements and argue for it as best we can.

7.2 The Deployment Scalability Tradeoff (DST)

>There is an inherent tradeoff between deployment scalability of a specification against being a simple, generic, general and resource limited.

We begin by noting again that this formulation fails to make explicit any of the assumptions that underlie the design of a spanning layer, namely the parameters $\prod$, N, v, and $\mathcal{E}$. Even if we suppress $\prod$, v, and $\mathcal{E}$ as unchanging underlying parameters, a less memorable but more complete formulation of the DST would be:

>There is an inherent tradeoff between deployment scalability of a specification of a layer that implements a particular set of necessary applications against being a simple, generic, general and resource limited.

The heart of the argument for the DST lies in the Hourglass Properties Theorem, which explains why a layer that is minimally sufficient for N will maximize the possible implementations of that layer. Having the maximum possible choice of implementations is a key element of deployment scalability. Being close enough to minimal sufficiency while maximizing the value of application set is the key implication of genericness.

**8. End-to-End Arguments Revisited**
One common formulation of the End-to-End Principle can be paraphrased as follows: "In a layered system, any feature should be located at the highest layer at which it can be correctly and completely implemented." While this formulation does not explain what the goal of this design rule is, we will assume for the purposes of this discussion that it is to create a layered system that is well adapted to achieving deployment scalability at all layers.

In the framework we have developed, this formulation can be interpreted as advocating that the spanning layer be minimally sufficient for the specified set of necessary applications (which is implicitly incorporated into the DST's notion of being "generic"). This interpretation does not make reference to simplicity, generality or resource limitation, but the argument for it mirrors our argument for the validity of the DST.

This statement of the End-to-End Principle is given in the imperative, and it does not address the question: "Or else what will happen?" The answer is sometimes given that failing to adhere to this principle will result in non-scalability of the resulting system. But scalability is not well-defined, this is difficult to make precise or to evaluate.

The answer to the question "Or else what?" that is suggested by the DST is that the more a system adheres to this principle the more deployment scalability it can potentially achieve. This statement is also not well-defined and is difficult to make precise or to evaluate, but an argument for it can be found in the Hourglass Properties of our formal model. Understood this way, strict rules against adding functionality to the spanning layer may be well-intentioned efforts to defend against loss of scalability, but they do not allow for a possible trade-off between the cost of diminished scalability and the value of increased functionality.

Some end-to-end arguments make more direct reference to simplicity, generality and resource limitation as elements of the desirable thinness of the waist of the hourglass. The DST is an attempt to join such reasoning with the formal framework that we have defined to explain the overall value of the hourglass model as a tool for achieving deployment scalability. If further development of the formal model were able to incorporate software engineering and economic considerations, then perhaps a more complete and rigorous account of the entire area will eventually be possible.

## 9. Examples and Applications (sketches)

Giving complete accounts of applications of the Deployment Scalability Tradeoff is a non-trivial matter, requiring the definition of the specification language, the program logic and its models, inferring predictions from the model and then arguing for the expected or experienced correctness of those predictions. I will sketch some possible applications and the anticipated results of such analysis here, and return to give more complete treatments later.

### 9.1 Fault Detection in TCP/IP

The classic example of the application of the End-to-End Principle, from which its name is derived, is the location of the detection of data corruption or packet loss or reordering in the TCP/IP stack [1]. One argument for the location of the detection of such faults at the endpoints of communication (historically perhaps the original argument) is that it cannot be completely accomplished hop-by-hop because this does not account for errors that occur *between hops*, in the mechanisms and functioning of the intermediate nodes (routers). Our account of the hourglass model does not account for this argument, but models a different one.

The scalability argument for end-to-end detection of faults is that removing such functions from the spanning layer makes it weaker, and therefore potentially admits more possible implementations. Because fault detection can be implemented above the spanning layer, the set of applications supported is not reduced. So one point about referring to our model is that it enables a clear separation of the basis for two quite different arguments regarding the placement of fault detection, which might have otherwise been conflated as comparable elements of the "thinness" of IP as the waist of the Internet hourglass.

Returning briefly to the argument that fault detection cannot be fully implemented hop-by-hop but can be implemented end-to-end, it is worth noting that it is less precise than the above scalability argument. In an end-to-end implementation of fault detection, there is still the possibility of error occurring within the implementation of TCP but outside the boundaries of the end-to-end check for errors. That is because sequence number and checksum verification occur within the mechanism of TCP, and there is some processing that occurs between those checks and the delivery of data to the application layer. Thus, while end-to-end checks reduce the locus of possible error from IP processing at every intermediate node plus all TCP processing to just a portion of the TCP processing at the endpoints, it does not in fact solve the problem completely in any formal or logical sense. I mention this difference not to disparage the effectiveness of TCP error detection, but simply to illustrate the difference between the scalability argument, which is based on formal logic, and the argument regarding the incompleteness of hop-by-hop checking, which is a matter of reducing the probability of error.

### 9.2 Process Creation in Unix

In early operating systems it was common for the creation of new processes to be a privileged operation that could be invoked only from code running with supervisory privileges. There were more than one reason for such caution, but one was that the allocation of operating system resources to create a new process was seen as too great to be delegated to the application level. Another reason was that the power of process definition (for example changing the identity under which the newly created process would run) was seen as too dangerous This led to a situation in which command line interpretation was a near-immutable function of the operating system that could only be changed by the installation of new supervisory code modules, often a privilege open only to the vendor or system administrator.

In Unix, process creation was reduced to the `fork()` operation, a logically much weaker operation that did not allow any of the attributes of the child process to be determined by the parent, but instead required that the child inherit such attributes from the parent [3]. Operations that changed sensitive properties of a process were factored out into orthogonal calls such as `chown()` and `nice()` which were fully or partially restricted to operating in supervisory mode, and `exec()` which was not but which was later extended with properties such as the setuid and sticky bits that were implemented as authenticated or protected features of the operating system. The decision was made to allow the allocation of kernel resources by applications, leaving open the possibility of dynamic management of such allocation by the kernel at runtime, and creating the possibility of "denial of service" type attacks that persists to this day.

The result of this design was not only the ability to implement a variety of command line interpreters as unpriviledged user processes, leading to innovations and the introduction of powerful new language features, but also the flexible use of `fork()` as a tool in the design of multitasking applications. This design approach has led to the adaptation of Unix and Unix-like kernels to highly varied user interfaces (such as mobile devices) that were not within the original Unix design space.

### 9.3 Data Replication and Placement in Logistical Networking

Network storage virtualization has become an important component of distributed information technology resource management systems. Data replication and placement is often incorporated as a feature of the storage spanning layer that defines community interoperability in such systems but which is not under the explicit management of clients of that layer, but accessible only through higher level abstractions. As a result, the policies that control such low level functions are either fixed or must be determined by clients through some policy interface of the virtualization layer.

The design of the Internet Backplane Protocol as the spanning layer of the Logistical Networking storage paradigm leaves the replication and placement of data to clients implementing higher layer functionality such as distributed file systems or content distribution networks [5]. Operations that allocate storage and store data to or move data between storage intermediate nodes (sometimes called a Storage Object Target but refered to in Logistical Networking parlance as a "depot") are local to the depot to which they are directed. To facilitate the implementation of dynamic data movement, direct third party data movement between network-adjacent depots is supported.

This design enables diverse policy mechanisms to be conveniently implemented by clients of the storage virtualization service without interference from possibly inappropriate policies (e.g. cache coherence) imposed in the implementation of the spanning layer. Clients that implement highly transient functions such as data streaming may decide to forgo replication, or to introduce it dynamically as a form of forward error correction only if network failure conditions are detected that indicate that it would be efficacious. Clients implementing more persistent functions such as content delivery might use replication and data distribution much more aggressively in order to localize data throughout the network and to maximize the profitability of diverse multipath data downloading algorithms by end users.

### 9.4 Grid Authentication

In a retrospective lecture on "tussle spaces" in the design of networks, I heard David Clark call out the lack of security at the Internet spanning layer as one regret. In today's difficult security environment, it is common to assume that some form of tight security is a necessity at the spanning layer, and in particular that authentication of identity should be a requirement of any use of common infrastructure.

The middleware framework for sharing of information technology resources that was given the communal name "The Grid" had strong authentication built in at the spanning layer of its protocol stack [4]. The Grid service stack was advertised as having a "thin waist" in analogy to the spanning layer of the Internet, and as an attempt to lay claim to the implication of scalability. Grid authentication required that ever user and resource under the management of the common middleware be assigned an X.509 Grid Certificate, obtainable only through a hierarchy of Certificate Authorities under the control of the U.S. Department of Energy or as similarly authoritative agency.

The exact impact of this requirement on the deployment scalability of the Grid is open to debate, but there is no question that a spanning layer that did not make this requirement for all access to common services would have had a weaker waist which would have had more possible implementations. The issue of what part, if any, of the substantial storage, networking and computing resources that were foreseen as being under the management of strong Grid authentication could have been responsibly accessed without such authentication, and what the implication might have been for the deployment scalability of Grid middleware, is beyond the scope of this discussion.

### 9.5 Process Management in PlanetLab

PlanetLab is a platform for the allocation and use of distributed information technology resources in the form of intermediate nodes running a modified Linux kernel. PlanetLab nodes located throughout the United States, Europe and in some other parts of the world [9]. The "spanning layer" of the distributed community of PlanetLab users consists of the shell command line, Internet, standard network services (eg scp) with some extensions for "slice" management and a Linux kernel modified to implement increased isolation of resource utilization between slices. Resources of the intermediate node are allocated by executing commands and running servers that service requests of their own client communities.

The NSF-sponsored Global Environment for Network Innovation (GENI) also had with ambitious plans to provide a scalable network virtualization platform, and succeeding in some of those goals. Today, the inheritors of the mantle of network diversity lie in Software Defined Networking and Network Function Virtualization. Perhaps the hourglass can provide an analytical tool to help predict the likelihood that these approach will actually scale in deployment if their functionality is implemented in the spanning layer of a network or distributed system.

### Acknowledgement

The contributors to this work have been many over the past two decades, and I will add an extensive list of grateful acknowledgments. However, I cannot release even the most preliminary version of this paper without acknowledging two people: Martin Swany, the "student" who first taught me about End-to-End, and my closest collaborator, Terry Moore, who has propelled my work and thinking through countless hours of challenging conversations and dialog. A philosopher by education and practice, Terry has more than once urged me to synthesize and publish this work by referring to Emerson's essay "Self-Reliance":

> "Speak your latent conviction, and it shall be universal sense; for always the inmost becomes the outmost,—— and our first thought is rendered back to us by the trumpets of the Last Judgment."

**Appendix A: Formal Definitions and Proofs (incomplete)**

> *"While precise reasoning is a revolutionary necessity, an obsession with formalism is a bourgeois disease." – Che Guevara?*

In this section we indicate the definitions of the fundamental concepts that we use as tools in this paper, namely service specifications, programs and their models, but stop short of fully formalizing these definitions or validating them in terms of model theory. We give proofs of the properties that we require in order to make our arguments in terms of the fundamental principles of model theory. We hope that these incomplete efforts will make the arguments of this paper plausible and that perhaps the definitions and results we seek can be cited or derived from known work in the area.

1.1 Service Specifications and the Implements Relation

1.1.1 We define a *service specification* to be a formal description of the syntax and necessary properties of a programming interface (API).

A service specification S is an API: it specifies the behavior of certain program elements (functions or subprograms) through statements expressed in a programming logic. For instance, these might be such statements:

    Forall A, B in Integer.  (A+1)+B = (A+B)+1
    Forall X, Y in Float.     (X > 1) Y := X*X  { Y > X }

In formal terms a service specification is a theory of the programming logic.

1.1.2 A *model* of a program is an environment in which all the undefined elements of the program are bound to corresponding objects (instantiated), such as functions or state transformers, depending on the programming logic. We can then talk about a model M satisfying a theory T (M ⊨ T), which means that it meets the API describe by T.

1.1.3 If P is a program, we define the *meaning* of P, or M[P], to be a model which corresponds to its *semantics*.

1.1.4 We define an *implements* relation $\prec_P$ between two service specifications S and T and a program P as follows:
- $T \prec_P S$ iff
- $(M[P] \vDash S) \Rightarrow (M[P] \vDash T)$

1.1.5 Discussion
The implements relation is intended to be analogous to the "reduces to" relation of structural complexity theory.

"In a model where API S is correctly instantiated, the program P correctly implements API T in terms of the instantiation of S."

1.2 Pre- and Postimages

We will express our formal analogs to scalability in terms of how large the classes of models are that can implement or can be implemented using a specification. We define pre- and post-image of a specification under implementation as follows:

1.2.1 $pre_\Pi(S) = \{T \mid \exists P \in \Pi.\ S \prec_P T\}$

1.2.2 $post_\Pi(S) = \{T \mid \exists P \in \Pi.\ T \prec_P S\}$

**2. Characterizing Application and Implementation Richness**

2.1 A specification S1
- is *more implementation rich* than another specification S2 iff
- $pre_\Pi(S1) \supseteq pre_\Pi(S2)$.

2.2 A specification S1
- is *more application rich* than another specification S2 iff
- $post_\Pi(S1) \supseteq post_\Pi(S2)$.

2.3 Weakness

A specification S1 is *weaker* than another specification S2 iff $S2 \vdash S1$. S1 is strictly weaker than S2 if it S1 is weaker than S2 but S2 is not weaker than S1.

**3. The Hourglass Properties**

If a specification S1 is weaker than another specification S2, then
1. S2 is more application rich than S1. and
2. S1 is more implementation rich than S2.

Proof:
1. For any specification T, if $S2 \prec_P T$, then
   - $S2 \vdash S1$ by the definition of weakness, so
   - $S1 \prec_P T$
   - $post_\Pi(S2) \supseteq post_\Pi(S1)$.

2. For any specification T, if $T \prec_P S1$ then
   - $S2 \vdash S1$ by the definition of weakness, so
   - $T \prec_P S2$, so
   - $pre_\Pi(S1) \supseteq pre_\Pi(S2)$.

**Appendix B: Some Pictures**

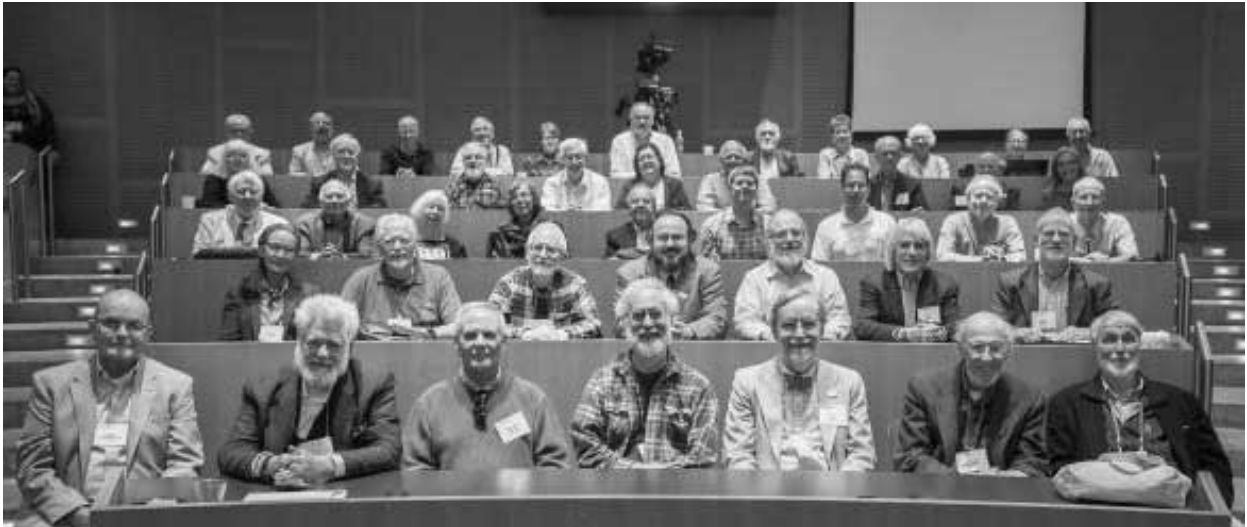
The Project MAC 50th Anniversary and Multics Reunion, 2014

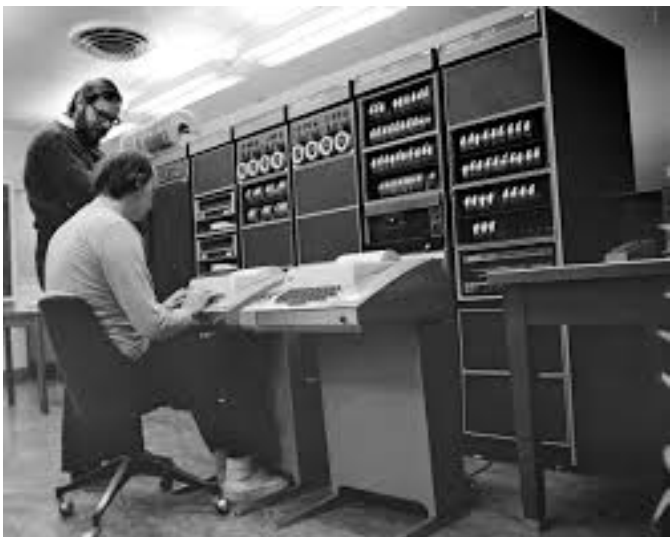
Ritchie and Thompson at Bell Laboratories, circa 1970s

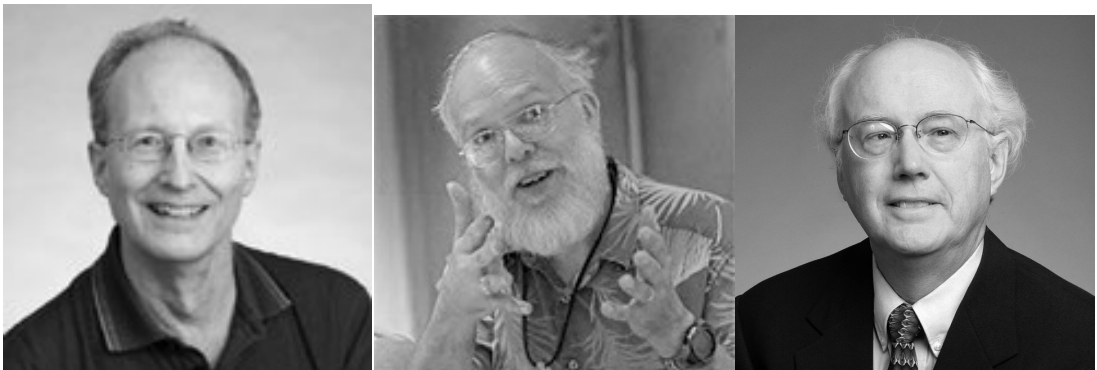
Saltzer, Reed and Clark